\documentclass{aa} 
\usepackage{txfonts}
\usepackage[latin1]{inputenc}  
\usepackage{graphicx}

\begin{document}
\title{Radiolysis of H$_2$O:CO$_2$ ices by heavy energetic cosmic ray analogs}
 \author{S. Pilling\inst{1}, E. Seperuelo Duarte\inst{2}, A. Domaracka\inst{3},
  H. Rothard\inst{3}, P. Boduch\inst{3}\and E.F. da Silveira\inst{4}}
 \institute{
 IP\&D/UNIVAP, Av. Shishima Hifumi, 2911, São Jose dos Campos, SP, Brazil.
 \and
 Grupo de Física e Astronomia, CEFET/Química de Nilópolis, Rua Lúcio Tavares, 1052,  CEP 2653-060, Nilópolis, Brazil.
 \and
  Centre de Recherche sur les Ions, les Matériaux et la Photonique (CEA /CNRS /ENSICAEN /Université de Caen-Basse Normandie), CIMAP - CIRIL - GANIL, Boulevard Henri Becquerel, BP 5133, F-14070 Caen Cedex 05, France.
 \and
 Departamento de Física, Pontifícia Universidade Católica do Rio de Janeiro
 (PUC-Rio), Rua Marquês de São Vicente, 225, CEP 22453-900, Rio de Janeiro, Brazil.
  }
\offprints{S. Pilling,\\ \email{sergiopilling@pq.cnpq.br}}
\date{Received / Accepted}
%
\abstract{An experimental study on the interaction of heavy, highly charged, and energetic ions (52 MeV $^{58}$Ni$^{13+}$) with pure H$_2$O, pure CO$_2$ and mixed H$_2$O:CO$_2$ astrophysical ice analogs is presented. This analysis aims to simulate the chemical and the physicochemical interactions induced by heavy cosmic rays inside dense and cold astrophysical environments such as molecular clouds or protostellar clouds. The measurements were performed at the heavy ion accelerator GANIL (Grand Accelerateur National d´Ions Lourds in Caen, France). The gas samples were deposited onto a CsI substrate at 13 K. \textit{In-situ} analysis was performed by a Fourier transform infrared (FTIR) spectrometer at different fluences. Radiolysis yields of the produced species were quantified.
The dissociation cross sections of pure H$_2$O and CO$_2$ ices are 1.1 and 1.9 $\times 10^{-13}$ cm$^2$, respectively. In the case of mixed H$_2$O:CO$_2$ (10:1) the dissociation cross sections of both species are about 1 $\times 10^{-13}$ cm$^2$. The measured sputtering yield of pure CO$_2$ ice is 2.2 $\times 10^4$ molec ion$^{-1}$. After a fluence of 2-3 $\times 10^{12}$ ions cm$^{-2}$ the CO$_2$/CO ratio becomes roughly constant ($\sim$0.1), independent of the of initial CO$_2$/H$_2$O ratio. A similar behavior is observed for the H$_2$O$_2$/H$_2$O ratio which stabilizes at 0.01, independent of the initial H$_2$O column density or relative abundance.

}

\titlerunning{Radiolysis of water-CO$_2$ ices by energetic heavy ions}
\authorrunning{Pilling et al.}
\maketitle

\section{Introduction}

H$_2$O and CO$_2$ are most abundant constituents of icy grain mantles in the interstellar medium (Whittet et al. 1996; Ehrenfreund \& Charnley 2000). Following Gerakines et al. 1999, a significant fraction of interstellar solid CO$_2$ exists in mixtures dominated by H$_2$O in both quiescent cloud and protostellar regions. In those regions the relative abundance of CO$_2$ with respect to H$_2$O ranges from 4\% to 25\%.

Inside the solar system the presence of H$_2$O and CO$_2$ ices is also ubiquitous. For example, CO$_2$ is widely detected in cometary ices. Its abundance, relative to H$_2$O, is about 6-7\% in the comets Hale-Bopp (Crovisier et al. 1997; Crovisier 1998) and Hyakutake (Bockelée-Morvan 1997). H$_2$O and CO$_2$ ice features have been observed also in the IR spectra of the icy Galilean satellites Europa, Ganymede, Callisto (Carlson et al. 1999; McCord et al. 1998), at Triton - the largest moon of Neptune (Quirico et al. 1999) - and at surface of Mars (Herr \& Pimentel 1969; Larson \& Fink 1972).

Deep inside dense molecular clouds and protostellar disks, as well as at the surfaces of solar system bodies surrounded by thick atmospheres, the frozen compounds are protected from stellar energetic UV photons. However, X-rays and energetic cosmic rays can penetrate deeper and trigger molecular dissociation, chemical reactions and evaporation processes. In the case of solar system ices without significant atmospheres such as Europa, Enceladus and Oort cloud comets, the material protection to UV photons is supplied by the upper layers (tens nm) of ice  and only the energetic particles reach the inner layers (e.g. 52 MeV Ni penetrate roughly 20 $\mu$m in pure water ice). These ices are directly exposed to stellar photons and comics ray, solar wind ($\sim$ 1 keV/amu ionized particles), energetic solar particles and/or planetary magnetosphere particles.

Laboratory studies and astronomical observations indicate that photolysis and radiolysis of such ices can create simple molecules such as CO, CO$_3$, O$_3$, H$_2$CO$_3$ and H$_2$O$_2$ and others more complex organic compounds such as formic acid, formaldehyde, methanol, etc. (e.g. Gerakines et al. 2000; Moore and Hudson 2000; Brucato et al. 1997, Wu et al. 2003). By comparing laboratory data (production rates, formation cross section and half-lives for these compounds) with astronomical observation, we will better understand the physicochemical processes occurring in the astronomical sources.

In this work, we present infrared measurements of two different mixtures of H$_2$O:CO$_2$ ices as well as of pure water and CO$_2$ ices, irradiated by 52 MeV $^{58}$Ni$^{13+}$. In section 2, the experimental setup is briefly described. The infrared spectra of frozen sample measured as a function of ion fluences are the subject of section 3. In section 4 a discussion is made about the dissociation cross sections and other radiolysis-induced effects, as well as on the formation of new species on ices. Astrophysical implication is discussed in section 5. Final remarks and conclusions are given in section 6.

\section{Experimental}

To simulate the effects of heavy and highly ionized cosmic rays on astrophysical cold surfaces, the facilities at the heavy ion accelerator GANIL (Grand Accelerateur National d'Ions Lourds, Caen, France) have been used. The gas samples (purity superior to 99\%) were deposited onto a CsI substrate at 13 K through a facing gas inlet. The sample-cryostat system can be rotated 180$^\circ$ and fixed at three different positions to allow: i) gas deposition; ii) FTIR measurement; and iii) perpendicular ion irradiation.

52 MeV $^{58}$Ni$^{13+}$ ion projectiles impinge perpendicularly onto H$_2$O:CO$_2$ ice (1:1), H$_2$O:CO$_2$ ice (10:1), as well as onto pure H$_2$O and CO$_2$ ices.
The target ionizing effects of Ni and Fe projectiles, at the same velocity, are almost identical since they have almost the same atomic number (Seperuelo Duarte et al. 2010). The incoming charge state $13+$ corresponds approximately to the equilibrium charge state after several collisions of 52 MeV Ni atoms (independent of the initial charge state) with matter (e.g. Nastasi et al. 1996).

\textit{In-situ} analysis was performed by Nicolet FTIR spectrometer (Magna 550) from 4000 to 600 cm$^{-1}$ with resolution of 1 cm$^{-1}$. The spectra were collected at different fluences up to $2 \times 10^{13}$ ions cm$^{-2}$. The Ni ion flux was $2 \times 10^{9}$ cm$^{-2}$ s$^{-1}$. During the experiment the chamber pressure was below $ 2\times 10^{-8}$ mbar due to the cryopumping by the thermal shield. Further details are given elsewhere (Seperuelo Duarte et al. 2009; Pilling et al. 2010).

Assuming an average density for the pure water ice and water-rich ices samples of 1 g/cm$^3$ and 1.8 g/cm$^3$ for the pure CO$_2$ ice, the thickness and the deposition rate are determined by using the initial molecular column density of the samples. For H$_2$O:CO$_2$ (1:1) and H$_2$O:CO$_2$ (10:1), the sample thicknesses were 0.6 and 0.4 $\mu$m. The deposition rates were roughly 7 and 3 $\mu$m/h, respectively. In the case of pure H$_2$O ice and pure CO$_2$ ice, the thicknesses were about 0.5 and 0.4 $\mu$m and, the deposition rates were 4 and 3 $\mu$m/h, respectively.  The penetration depth of 52 MeV Ni ions is much larger than ice thicknesses therefore the ions pass though the target with approximately the same velocity (cross sections are constant).

The molecular column density of a sample was determined from the relation between optical depth $\tau_\nu = \ln(I_0/I)$ and the band strength, A (cm molec$^{-1}$), of the respective sample  vibrational mode (see d´Hendecourt \& Allamandola 1986). In this expression, $I$ and  $I_0$ is the intensity of light at a specific frequency before and after passing through a sample, respectively. Since the absorbance measured by the FTIR spectrometer was $Abs_\nu=\log(I_0/I)$, the molecular column density of ice samples was given by
\begin{equation} \label{eq-N}
N= \frac{1}{A} \int \tau_\nu d\nu = \frac{2.3}{A} \int Abs_\nu d\nu \quad \textrm{[molec cm$^{-2}$],}
\end{equation}
where $Abs_\nu = \ln(I_0/I)/\ln(10) = \tau_\nu/2.3$.

The vibrational band positions and their infrared absorption coefficients (band strengths) used in this work are given in Table~\ref{tab:A}.

\begin{table}[!htb]
\caption{Infrared absorption coefficients (band strengths) used in the column density calculations for the observed molecules.} \label{tab:A}
\setlength{\tabcolsep}{3pt}
\begin{tabular}{ l l l l r }
\hline \hline
Frequency  & Wavelength & Assignment  & Band strength (A)                & Ref. \\
(cm$^{-1}$) & ($\mu$m)  &      &  (cm molec$^{-1}$)    &      \\
\hline
2342         & 4.27 & CO$_2$ ($\nu_3$)      & 7.6 $\times 10^{-17}$            & [1]  \\
3250   & 3.07 & H$_2$O ($\nu_1$)      & 2.0 $\times 10^{-16}$            & [1]  \\
$\sim$ 2850  & $\sim$3.5 & H$_2$O$_2$ ($\nu_2 + \nu_6$)  & 5.7 $\times 10^{-17}$     & [3]  \\
2139         & 4.67 & CO ($\nu_1$)                & 1.1 $\times 10^{-17}$       & [2]  \\
2044  & 4.89  & CO$_3$ ($\nu_3$)         & 8.9 $\times 10^{-17}$          & [4]  \\
1307         & 7.65  & H$_2$CO$_3$ (\tiny C-OH bend \footnotesize)  & 1.0 $\times 10^{-16}$     & [5]  \\
$\sim$ 1040  & $\sim$9.6  & O$_3$ ($\nu_3$)          & 1.4 $\times 10^{-17}$          & [6]  \\
\hline \hline\\
\end{tabular}
[1] Gerakines et al. 1995; [2] Gibb et al. 2000; [3] Loeffler et al. 2006; [4] Bennett et al. 2004 [5] Gerakines et al. 2000; [6] Smith et al. 1985.
\end{table}

\section{Results}

\begin{figure}[!ht]
 \centering
 \includegraphics[scale=0.50]{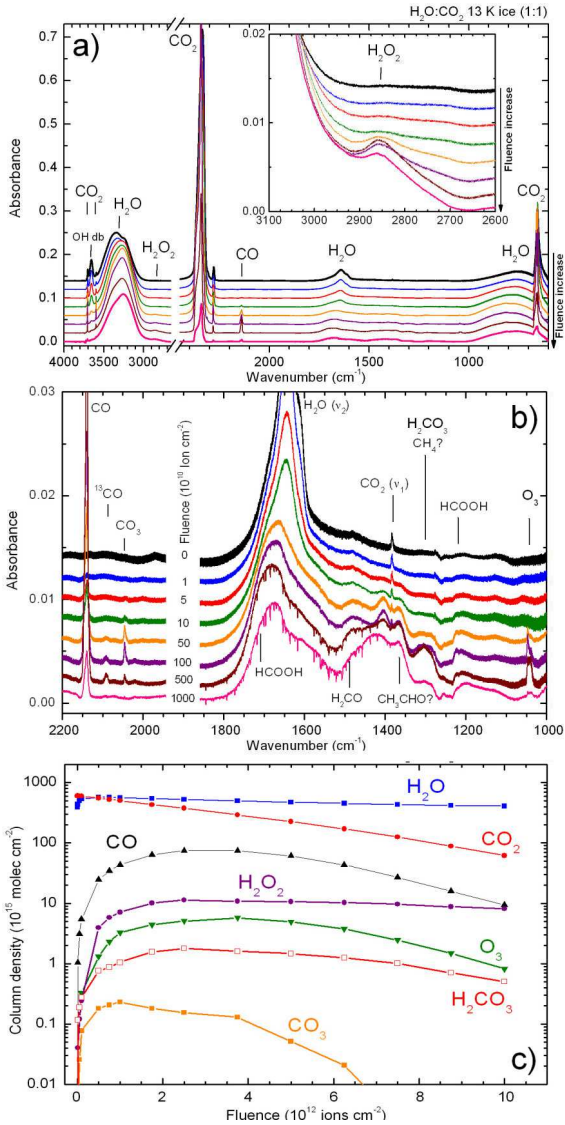}
\caption{a) Infrared spectra of H$_2$O:CO$_2$ 13 K ice (1:1) before (top dark line) and after different irradiation fluences. b) Expanded view from 2200 to 1000 cm$^{-1}$. c) Molecular column densities derived from the infrared spectra during the experiment. The lines are to guide the eyes.} \label{fig:EXP1-FTIR}
\end{figure}

Figure~\ref{fig:EXP1-FTIR}a presents the infrared spectra of H$_2$O:CO$_2$ ice (1:1) at 13 K, before (top trace) and after different irradiation fluences. Each spectrum has been shifted for clarity. The narrow peak at 2342 cm$^{-1}$ is the CO$_2$ stretching mode ($\nu_3$). The broad structures around 3300 and 800 cm$^{-1}$ correspond to the vibration modes of water, $\nu_1$ and $\nu_L$, respectively. The band at 1600 cm$^{-1}$ is due to the water $\nu_2$ vibration mode. The presence of the OH dangling band at about 3650 cm$^{-1}$ is also observed in the figure indicating a high porosity (Palumbo et al. 2006). The inset figure shows the newly formed H$_2$O$_2$ species due to the radiolysis of water inside the ice.

The region between 2200 to 1000 cm$^{-1}$ is shown in detail in Fig.~\ref{fig:EXP1-FTIR}b. The ion fluence of each spectra is also given. In this spectral region several peaks grow as a function of fluence: they correspond to the appearance of new species formed by the 52 MeV Ni atoms bombardment, including CO, CO$_3$, O$_3$, H$_2$CO$_3$, H$_2$CO (tentative), HCOOH (tentative). Some IR features originated from unidentified organic species appear around 1350 and 1450 cm$^{-1}$.

\begin{figure}[!t]
\centering
\includegraphics[scale=0.5]{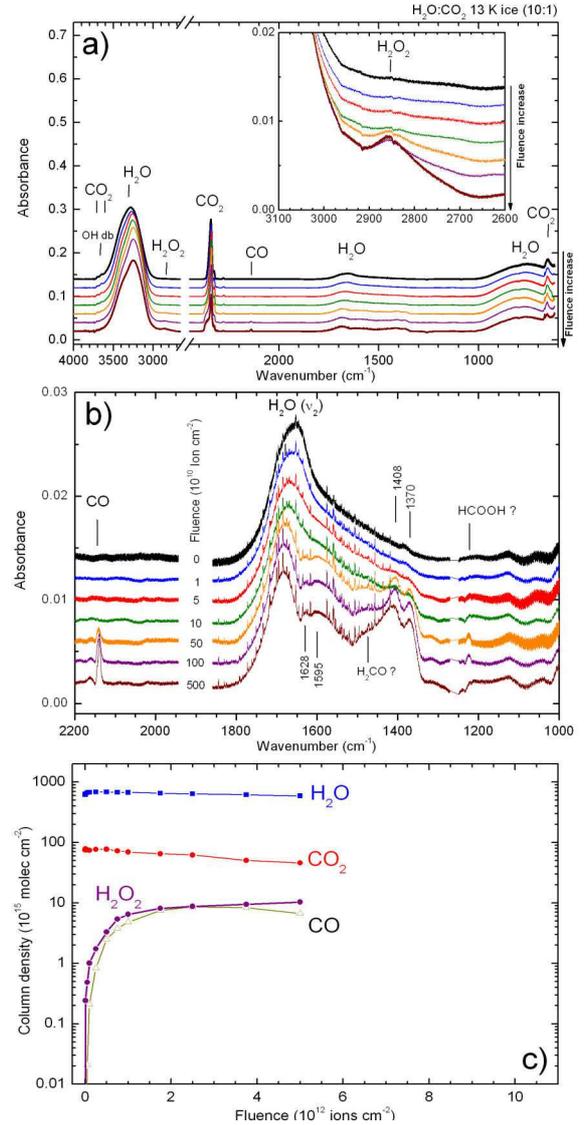}
\caption{a) Infrared spectra of H$_2$O:CO$_2$ 13 K ice (10:1) before (top dark line) and after different irradiation fluences. b) Expanded view from 2200 to 1000 cm$^{-1}$. c) Molecular column densities derived from the infrared spectra during the experiment. The lines are to guide the eyes.} \label{fig:EXP2-FTIR}
\end{figure}

The evolution of the column density as function of fluence is shown in  Fig.~\ref{fig:EXP1-FTIR}c. The infrared band strengths used for the determination of the column density are listed in Table~\ref{tab:A}. The column density of water presents constant concentration of about $5\times 10^{17}$ molecules cm$^{-2}$, decreasing very slowly as the fluence increases. This fact is attributed to an approximate compensation between persisted water condensation (layering) and its disappearance by sputtering and dissociation.

The initial enhancement in the water column density maybe associated with the compaction effect produced by ion bombardment as discussed elsewhere (Palumbo 2006, Pilling et al. 2010). This compaction changes the band strength of some vibrational modes. Leto and Baratta (2003) showed that the band strength of water  $\nu_1$ vibrational mode undergoes a 10\% increase during the first ion impacts (low fluence) on ice in experiments employing ions with energy of dozens of keV. Another explanation may be a strong water layering just on the begging of irradiation. The CO$_2$ abundance reaches its half value at a fluence of about $4\times 10^{12}$ ions cm$^{-2}$. The CO abundance increases very rapidly, reaching a maximum around 3-4$\times 10^{12}$ ions cm$^{-2}$ and decreasing after that. The same behavior was observed for the O$_3$, H$_2$CO$_3$ and CO$_3$, all produced by the radiolysis of CO$_2$ inside the ice. For O$_3$, H$_2$CO$_3$ and CO$_3$, the column density maximum occurs at 4 $\times 10^{12}$, 3 $\times 10^{12}$ and 1 $\times 10^{12}$ ions cm$^{-2}$, respectively. In the case of H$_2$O$_2$ the maximum occurs at around 1 $\times 10^{12}$ ions cm$^{-2}$ and remains constant ($\sim$ 1 $\times 10^{16}$ molec cm$^{-2}$) until the end of irradiation (up to a fluence of 1 $\times 10^{13}$ ions cm$^{-2}$).

\begin{figure}[!t]
 \centering
 \includegraphics[scale=0.47]{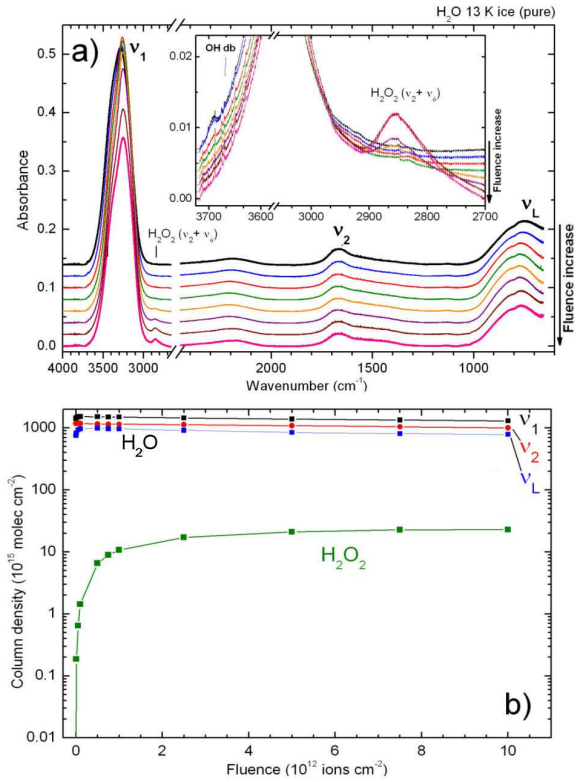}
 \caption{a) Infrared spectra of pure water ice at 13 K obtained for different 52 MeV Ni fluences. Inset figures present details of newly formed species. b) Molecular column densities derived from the infrared spectra during the experiment.} \label{fig:4-pureH2O}
\end{figure}
\begin{figure}[!t]
 \centering
 \includegraphics[scale=0.51]{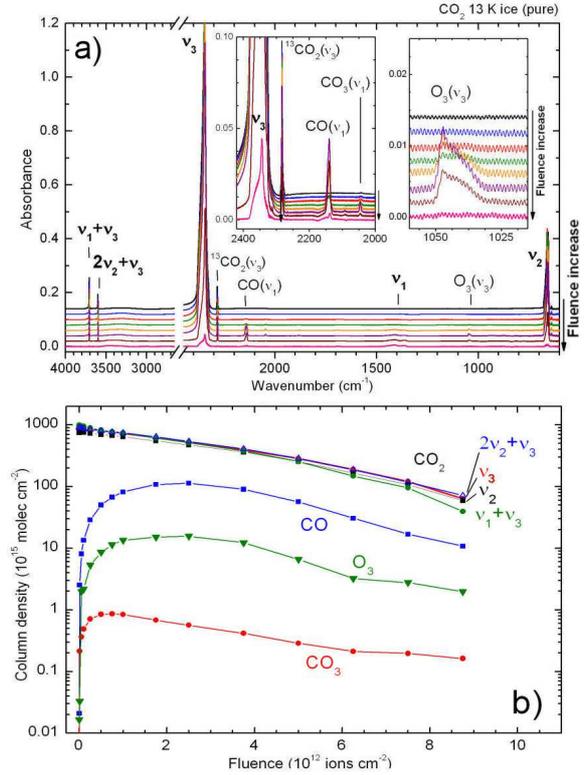}
 \caption{a) Infrared spectra of pure CO$_2$ ice at 13 K after different 52 MeV Ni  fluences. Inset figures present details of newly formed species. b) Molecular column densities derived from the infrared spectra during the experiment.} \label{fig:5-pureCO2}
\end{figure}

The infrared spectra of H$_2$O:CO$_2$ (10:1) ice before and after different irradiation fluences up to 5$\times 10^{12}$ ions cm$^{-2}$ are given in Fig.~\ref{fig:EXP2-FTIR}a. In contrast to H$_2$O:CO$_2$ (1:1), this ice does not present the OH dangling band at about 3650 cm$^{-1}$, may be due to the lower deposition rate or due to the low amount of CO$_2$ on ice. As suggested by Bouwman et al. (2007), the presence of impurities like CO and CO$_2$ on amorphous water ices increases the ice´s porosity and the presence of the OH db become more prominent.

Figure~\ref{fig:EXP2-FTIR}b shows details of the infrared spectra for regions between 2200 to 1000~cm$^{-1}$. The wavenumber of some unidentified IR frequencies are given. The infrared peaks tentatively attributed to H$_2$CO ($\sim$ 1470 cm$^{-1}$) and HCOOH ($\sim$ 1220 cm$^{-1}$) seem to be smaller than in the case of H$_2$O:CO$_2$ (10:1) ice. As in the previous case the appearance of the CO $\nu_1$ line at 2139 cm$^{-1}$ is due to the CO$_2$ radiolysis.

The variations of the column densities of the most abundant molecules observed during the irradiation of H$_2$O:CO$_2$ ice (10:1) by 52 MeV Ni ions as function of fluence are given in Fig.~\ref{fig:EXP2-FTIR}c. Peaks due to O$_3$, H$_2$CO$_3$ or CO$_3$ molecules were not observed.

The infrared spectra of pure H$_2$O and CO$_2$ ices at 13 K for different 52 MeV Ni fluences are shown in Fig.~\ref{fig:4-pureH2O}a and Fig.~\ref{fig:5-pureCO2}a, respectively. Upper curves indicate the virgin ice and figure insets display peak details of the newly formed species. In the case of the radiolysis of pure water, only the formation of H$_2$O$_2$ (vibration mode $\nu_2$+$\nu_6$) is observed. In contrast, after the radiolysis of CO$_2$ pure ice, at least three different compounds are formed: CO ($\nu_1$), O$_3$ ($\nu_3$) and CO$_3$ ($\nu_1$) (see also Seperuelo Duarte et al. 2009).

Figure~\ref{fig:4-pureH2O}b shows the evolution of water molecular column density derived from different vibration modes ($\nu_1$, $\nu_2$ and $\nu_L$) and the column density of H$_2$O$_2$ as a function of ion fluence. The band strengths employed for water vibration modes $\nu_2$ (1650 cm$^{-1}$) and $\nu_L$ ($\sim$ 800 cm$^{-1}$) were obtained by Gerakines et al (1995) and Gibb et al. (2000), respectively. The evolution of CO$_2$ column density, derived from different vibration modes ($\nu_2$, $\nu_3$, $\nu_1 + \nu_3$ and  $2\nu_2 + \nu_3$), is given in Fig.~\ref{fig:5-pureCO2}b: the distinct result are practically the same, as expected. The band strengths employed for CO$_2$ vibration modes $\nu_2$ (670 cm$^{-1}$), $2\nu_2 + \nu_3 $ (3599 cm$^{-1}$), and $\nu_1 + \nu_3 $ (3708 cm$^{-1}$), were obtained by Gerakines et al. (1995). The column densities of the new species produced by radiolysis are also shown and will be discussed further.

\begin{figure}[!t]
 \centering
 \includegraphics[scale=0.5]{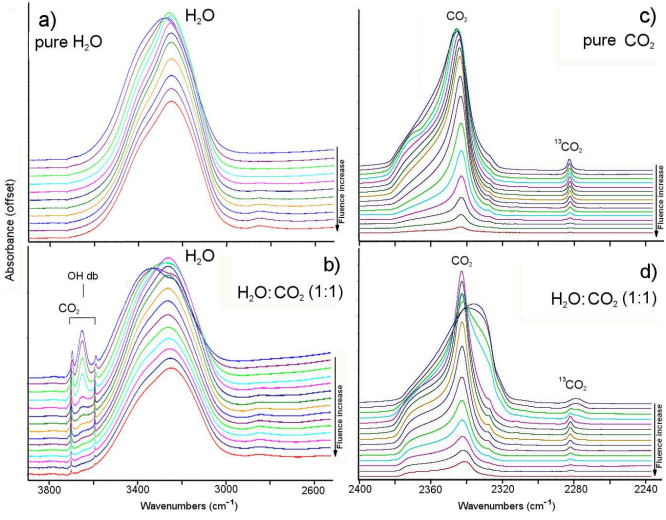}
 \caption{Selected profile of water and CO$_2$ vibration mode at difference fluences up to $10^{13}$ ions cm$^{-2}$. Pure water $\nu_1$ vibration mode (a), water $\nu_1$ mode in mixed ices (b), pure CO$_2$ $\nu _3$  mode (c) and CO$_2$ $\nu _3$ mode in mixed ices (d). See details in text.} \label{fig:ComparisonIR}
\end{figure}

The infrared absorption profile of water ($\nu_1$) and carbon dioxide ($\nu_3$) in pure and mixed (H$_2$O:CO$_2$ (1:1)) ices as a function of ion fluence are shown in Fig.~\ref{fig:ComparisonIR}. Upper panels (a and c) indicate pure ices. In the case of mixed ice, the selected IR features are illustrated in the bottom panels (b and d). In all panels the upper curves indicate the non irradiated spectrum. Each spectrum has a small offset for clearer visualization. The bombardment by heavy ions slightly shifts the water $\nu_1$ vibration mode to low frequencies. Since different water clusters are responsible for this IR band (Paul et al. 1997) this suggests that small water clusters within the ice are being converted to larger cluster structures. Also large clusters may be less radiation sensitive than small clusters. This behavior is enhanced in the case of mixed ices (Fig.~\ref{fig:ComparisonIR}b).

Figure~\ref{fig:ComparisonIR}b presents the OH dangling bands ($\sim$3670~cm$^{-1}$) attributed to water molecules weakly adsorbed into micropores inside the ice (see Palumbo 2006). As the fluence increases, the micropores collapse and the OH dangling vibration modes decrease. We have shown previously (Pilling et al. 2010) that the ice compaction produced by heavy cosmic rays are at least 3 orders of magnitude higher than that produced by (0.8~MeV) protons.

The profile of the $\nu_3$ vibration mode of carbon dioxide appears to be very sensitive to the presence of water. It becomes very broad and also presents a shift toward the lower frequencies when water is mixed with CO$_2$ (Fig.~\ref{fig:ComparisonIR}d). As the ion fluence increases, the $\nu_3$ band of CO$_2$ becomes sharp and the peak moves toward higher frequencies becoming similar to pure CO$_2$ ice. A detailed discussion on the effects of CO$_2$ and H$_2$O on band profiles observed in mixed H$_2$O:CO$_2$ ices is given elsewhere (Öberg et al. 2007).

\section{Discussion}

\subsection{Dissociation cross section}

As discussed previously by Seperuelo Duarte et al. (2009) and Pilling et al. (2010), the complex interaction between an energetic heavy ion with an ice target may be described through partial scenarios by considering successive aspects of the phenomenon. Since the impact of 52 MeV Ni atoms lies in electronic energy loss domain (projectile velocity $\gtrsim$ 0.01 cm $\mu$s$^{-1}$) most of the deposited energy leads to excitation/ionization of target electrons (electronic energy loss regime). In turn, the electrons liberated from the ion track core (0.3 nm of diameter for 52 MeV Ni atoms; Iza et al. 2006), transfer their energy to the surrounding condensed molecules ($\sim$ 3 nm). The re-neutralization of the track proceeds concomitantly with the local temperature rise, leading to a possible sublimation.

As discussed previously (Pilling et al 2010), the column density rate (a function of fluence) for each condensed molecular species during radiolysis can be written as:
\begin{equation} \label{eq:PAI}
\frac{dN_i}{dF} = \sum_{j \neq i} \sigma_{f,ij} N_j + L_i  - \sigma_{d,i} N_i - Y_i \Omega_i(F)
\end{equation}
where $\sum_j \sigma_{f,ij} N_j$ represents the total molecular production rate of the $i$ species directly from the $j$ species, $L_i$ is the layering, $\sigma_{d,i}$ is the dissociation cross section, $Y_i$ is the sputtering yield of and $\Omega_i(F)$ is the relative area occupied by the $i$ species on the ice surface.

Considering that the analyzed compounds cannot react in one-step process to form another species originally present in ice (i.e., $\sigma_{f,ij}\approx 0$), in the case of pure ices Eq.~\ref{eq:PAI} can be written by:

\begin{equation}\label{eq:N1}
N_i= (N_{0,i} - N_{\infty,i}) \exp(-\sigma_{d,i} F) + N_{\infty,i}
\end{equation}
where $N_{\infty,i} = (L_i - Y_i)/ \sigma_{d,i}$ is the asymptotic value of column density at higher fluences due to the presence of layering. $N_i$ and $N_{0,i}$ are the column density of species $i$ at a given fluence and at the beginning of experiment, respectively. In the case of pure $CO_2$ the layering yield is negligible $L_{CO2}=0$.

In the case of water-mixed ices, after the rapid ice compaction phase at the beginning of the irradiation, water layering tends to recover the ice surface, $\Omega_i(F) \rightarrow \delta_{i1}$, preventing progressively sputtering of CO$_2$ molecules. At the end of these processes, the system of Eq.~\ref{eq:PAI} is decoupled and can be solved analytically by:

\begin{equation}\label{eq:NH2O}
N= (N_{0} - N_{\infty}) \exp(-\sigma_{d} F) + N_{\infty} \quad \textrm{for H$_2$O}
\end{equation}
and

\begin{equation} \label{eq:NCO2}
N = N_{0} \exp(-\sigma_{d} F)  \quad \textrm{for CO$_2$}
\end{equation}
where $N_{\infty} = (L - Y)/ \sigma_{d}$ is the asymptotic value of column density of water due to the presence of layering.

\begin{table*}[!htb]
\begin{center} %
\caption{Dissociation cross section, radiochemical yield and sputtering values of H$_2$O and CO$_2$ obtained from the radiolysis of pure water ice and mixed water-CO$_2$ ices by 52 MeV Ni ions. } \label{tab:SIG-Destruct}
\setlength{\tabcolsep}{3pt}
\begin{tabular}{ l l c c c c c c c }
\hline \hline
Species  & Mixture & $\sigma_d$  & -G$_d$ & $N_\infty$  & $Y$   &  $L^\ddag$ &  $N_0$   & $Model$    \\
 & (H$_2$O:CO$_2$) & $(10^{-13}$ cm$^2$) & molec /100 eV  & (10$^{17}$ molec cm$^{-2}$)  & (10$^4$ molec ion$^{-1}$)  & (10$^4$molec ion$^{-1}$) & (10$^{17}$ molec cm$^{-2}$)   &    \\
\hline
H$_2$O      & (1:0)    & 1.1     & 7.48 & 12        & 1$^a$    & 14       & 15      & 1 \\
            & (10:1)   & $\sim$1 & 4.90 & $\sim$ 4  & 1$^a$    & $\sim$5  & 6.9     & 2 \\
            & (1:1)    & 1.0     & 6.35 & 3.0       & 1$^a$    & 4.0      & 5.9     & 3 \\
CO$_2$      & (0:1)    & 1.8     & 6.90 & NA$^\dag$ & 2.2      & 0        & 9.1     & 4\\
            & (1:1)    & 1.6     & 7.84 & NA        & 1.2$^b$   & 0        & 6.2     & 5a \\
            & (1:1)    & 2.1     & 10.3 & NA        & 0$^c$    & 0      & 6.2        & 5b \\
            & (10:1) & $\sim$1   & 6.35 & NA        & $\sim$0.2$^d$   & 0      & $\sim$0.8  & 6a\\
            & (10:1) & $\sim$0.7 & 4.45 & NA        & 0$^c$       & 0      & $\sim$0.8  & 6b\\
\hline \hline
\end{tabular}
\end{center}
$^a$ Taken from Brown et al. (1984).
$^b$ Estimated to be a half of the value determined from pure CO$_2$ ice.
$^c$ No CO$_2$ sputtering. Assuming that the water layering is high enough to fully cover the CO$_2$ molecules on the surface.
$^d$ Estimated to be a tenth of the value determined from pure CO$_2$ ice.
$^\dag$ NA=Non applied.
$^\ddag$ For H$_2$O: Derived from $L = N_\infty \sigma_{d}+Y$. For CO$_2$: Assuming no layering due to residual CO$_2$.
\end{table*}

As discussed by Loeffler et al. (2005), the radiochemical formation yield ($G_f$) of a given compound per 100 eV of deposited energy, at normal incidence, can be expressed in terms of the formation cross section ($\sigma_f$) and the stopping power (S), in units of eV cm$^2$/molec, as:

\begin{equation} \label{radiochemical}
G_f = 100 \frac{\sigma_f}{S}   \quad \textrm{molecule per 100 eV}
\end{equation}

This definition can be extended for radiochemical dissociation (destruction) yield of a given compound per 100 eV of deposited energy ($G_d$) by replacing the formation cross section in Eq.~\ref{radiochemical} by the negative value of the dissociation cross section (-$\sigma_d$). Therefore, negative $G_d$ values indicate that molecules are being dissociated or destroyed after energy deposition into the ice.

Adopting the $S$ values from the Stopping and Ranges module of the SRIM2003 package (Ziegler 2003), the values of the radiation yield $G$ in the experiments can be determined and compared with literature values. Table 2 presents the radiochemical destruction yields of H$_2$O and CO$_2$ per 100 eV of deposited energy in each calculated model. The stopping powers of 52 MeV Ni ions for pure water ice and for CO$_2$ ice are $S = 1.4 \times 10^{-12}$ eV cm$^2$/H$_2$O and 2.6 $\times 10^{-12}$ eV cm$^2$/CO$_2$, respectively. In the case of mixed H$_2$O:CO$_2$ ices the stopping power is determined by interpolation between the pure ice values. We considered that the chemical changes during the irradiation does not affect the stopping power.

\begin{figure}[!hb]
 \centering
\includegraphics[scale=0.62]{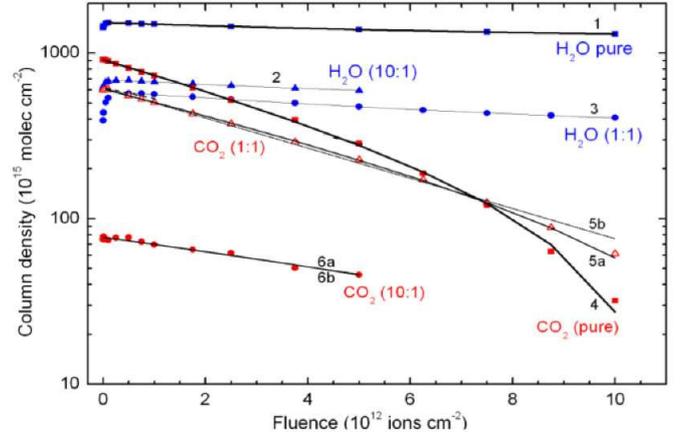}
\caption{Variation of the experimental column densities of H$_2$O and CO$_2$ as a function of fluence. Lines represent the fittings using Eq.~\ref{eq:N1} for pure ices and  Eqs.~\ref{eq:NH2O} and \ref{eq:NCO2} for mixed ices. The parameters models are listed in Table~\ref{tab:SIG-Destruct}.} \label{fig:fitting}
\end{figure}

Figure~\ref{fig:fitting} presents the fitting curves for the H$_2$O and CO$_2$ column densities employing Eq.~\ref{eq:N1} (for pure ices), and Eqs.~\ref{eq:NH2O} and \ref{eq:NCO2} (for mixed ices). Numeric labels indicate parameters listed in Table~\ref{tab:SIG-Destruct}. The H$_2$O column density data level off around $ N_\infty \simeq 1.2 \times 10^{18}$, $4 \times 10^{17}$, and $3 \times 10^{17}$ molec cm$^{-2}$ for pure water, H$_2$O:CO$_2$ (10:1) and H$_2$O:CO$_2$ (1:1) ice samples, respectively.
The average value for the dissociation cross section of water, employing Eq.~\ref{eq:N1}, is $\sigma_{d} \sim 1 \times 10^{-13}$ cm$^2$. The sputtering yield for water measured by Brown et al. (1984) was extrapolated for the 52 MeV Ni ions impact as discussed in Pilling et al. (2010). The obtained value is $Y = 1 \times 10^4$ molecules per impact. The estimated average water layering of both experiments was within the 4-14$\times 10^{4}$ molec ion$^{-1}$ range, obtained from the relation $L = N_\infty \sigma_{d}+Y$. Due to the large value of water layering, CO$_2$ species in the mixed ices were recovered by a H$_2$O film and the CO$_2$ ice sputtering yield is considered negligible in the current experiments; therefore, the data are adjusted directly by Eq.~\ref{eq:NCO2}.

The fitting parameters (dissociation cross section, radiochemical yield, sputtering, layering and the initial relative molecular abundance of each species in the ices) are listed in Table~\ref{tab:SIG-Destruct}. Pure H$_2$O and CO$_2$ ice were fitted by the parameters of models 1 and 4, respectively. For water species the layering yield was determined by $L = N\inf + \sigma_d + Y$. For CO$_2$ species (model 4 to 6b), it is assumed no layering due to residual CO$_2$. Model 5a is the fitting for the CO$_2$ species in the mixed H$_2$O:CO$_2$ ice (1:1) and the sputtered value was considered to be half the value from pure CO$_2$ ice.

In model 5b it is assumed that the water layering is high enough to fully cover the CO$_2$ on the surface producing a negligible sputtering yield of CO$_2$ molecules. From the comparison of models 5a and 5b we observe that the error in the dissociation cross section is lower than 30\%. Model 6a concerns the fitting for the CO$_2$ species in the mixed H$_2$O:CO$_2$ (10:1) ice; the sputtering yield is considered to be a tenth of the value determined from pure CO$_2$ ice.

From models 5 we observe that the precise determination of sputtering is possible only for the ices monitored at higher ion fluences ($\sim10^{13}$ ion cm$^{-2}$) since different sets between sputtering and dissociation cross section can be adjusted for the data with low fluences ($< 5 \times 10^{12}$ ions cm$^{-2}$). Due to the degeneracy of such model for low ion fluence IR data, the error observed in the dissociation cross section on models 6 is about 40\%. Table~\ref{tab:SIG-Destruct} shows that the dissociation cross section at typical astrophysical ice ([CO$_2$]/[H$_2$O] $\sim$ 0.05-0.1) are similar and equal to $\sim 1 \times 10^{-13}$ cm$^{-2}$.

\subsection{Synthesis of new species}

\begin{table*}[!htb]
\begin{center} %
\caption{Formation and dissociation cross section of newly formed species from the radiolysis of H$_2$O:CO$_2$ ices by 52 MeV Ni ions.} \label{tab:SIG-formation}
\setlength{\tabcolsep}{5pt}
\begin{tabular}{ l l c c c c c c }
\hline \hline
Species  & Parental      & Mixture           & $\sigma_f$           & $\sigma_d$        & $G_f$  & $-G_d$  & Model \\
         & species       & [H$_2$O:CO$_2$]   & $(10^{-13}$ cm$^2$)  & $(10^{-13}$ cm$^2$) & (molec /100 eV)     &  (molec /100 eV) &       \\
\hline
H$_2$O$_2$ & H$_2$O           & [1:0]   & 0.19  & 10   & 1.3  & 68.0  & 1\\
           & H$_2$O           & [10:1]  & 0.13  & 9.3  & 0.87 & 59.1  & 2 \\
           & H$_2$O           & [1:1]   & 0.10  & 6.9  & 0.48 & 33.8  & 3 \\
CO        & CO$_2$            & [0:1]   & 1.8   & 11   & 6.8  & 42.1  & 4\\
          & CO$_2$            & [1:1]   & 1.2   & 7.3  & 5.8  & 35.8  & 5\\
          & CO$_2$            & [10:1]  & 0.91  & 4.4  & 5.8  & 27.9  & 6\\
O$_3$     & CO$_2$            & [0:1]   & 0.27  & 16   & 1.1  & 61.3  & 7\\
          & CO$_2$            & [1:1]   & 0.06  & 5.0  & 0.30 & 24.5  & 8\\
CO$_3$    & CO$_2$            & [0:1]   & 0.09  & 97   & 0.33 & 372   & 9 \\
          & CO$_2$            & [1:1]   & 0.01  & 31   & 0.06 & 152   & 10\\
H$_2$CO$_3$ & H$_2$O and CO$_2$ & [1:1] & 0.03  & 9.9  & 0.14 & 48.6  & 11\\
\hline \hline
\end{tabular}
\end{center}
\end{table*}

\begin{figure}[!h b]
 \centering
\includegraphics[scale=0.6]{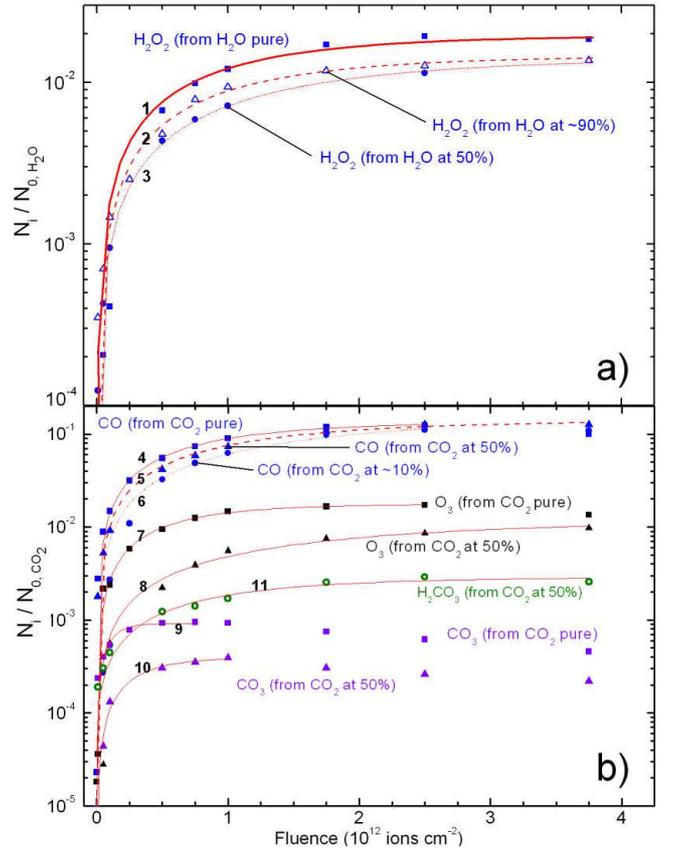}
 \caption{The evolution of newly formed species from the bombardments of H$_2$O:CO$_2$ ices by 52 MeV Ni as a function of fluence. a) Production of H$_2$O$_2$ from H$_2$O molecules in different ices. b) Production of CO, O$_3$,  CO$_3$ and H$_2$CO$_3$ from CO$_2$ molecules in different ices. The relative abundance of each parental species are indicated. The lines indicate the fittings using Eq.~\ref{eq:SigmaNEW} and the model parameters are given in Table~\ref{tab:SIG-formation}.} \label{fig:NiNp}
\end{figure}

The evolution of the column density of the newly formed species from H$_2$O:CO$_2$ ices as a function of fluences is shown in Fig.~\ref{fig:NiNp}a and Fig.~\ref{fig:NiNp}b. The ratio of H$_2$O$_2$ column density over its parent molecule (H$_2$O) initial column density, $N_{H_2O_2}/N_{0,H_2O}$, as a function of fluence in 3 different water-concentration ices (pure H$_2$O ice, $\sim$50\% and $\sim$90\%) is given in Fig.~\ref{fig:NiNp}a. The evolution of newly formed species CO, O$_3$, and H$_2$CO$_3$ from CO$_2$-rich ices as a function of fluence is presented in Fig.~\ref{fig:NiNp}b. $N_i/N_{0,CO_2}$ indicates the column density ratio of a given produced species over its parent molecule (CO$_2$) initial column density. Three different CO$_2$-rich ices are analyzed (pure CO$_2$ ice, $\sim$50\% and $\sim$10\%). The data was fitted by the simple exponential associate expression:

\begin{equation}\label{eq:SigmaForm}
N_i/N_{0,p}= A_i + B_i [1- \exp(-\sigma_{d,i} F)]
\end{equation}

where $N_i$ and $N_{0,p}$ represents the column density of a given produced species $i$ over its parent molecule $p$ (H$_2$O or CO$_2$). A$_i$ and B$_i$ are constants which indicate the initial and the maximum amount (after radiolysis) of species $i$ on the ice, $\sigma_{f,i}$ is the formation cross section and $F$ is the ion fluence.

As the fluence increases, the number of produced molecules also increases and reaches a maximum. At this moment, the number of molecules produced by radiolysis of parental species is equal to the number of new molecules that dissociate possibly in the reverse reaction set (e.g. $p \rightleftarrows i + j + k + $...).  At the equilibrium the column density of the parental and daughter species, $N_{eq,p}$ and $N_{eq,i}$, respectively, are related by the expression:
\begin{equation}\label{eq:SigmaF2}
N_{eq,p} ~ \sigma_{f,i} = N_{eq,i} ~ \sigma_{d,i}
\end{equation}
where $\sigma_{f,i}$ and $\sigma_{d,i}$ is the formation cross section and dissociation cross section of daughter species $i$.

Since the column densities of daughters species $i$ are much smaller than the one observed for parental species $p$, as a first approximation we have the following relation between them:
\begin{equation}\label{eq:SIG2}
\frac{N_{eq,i}} { N_{0,p}}  + \frac{N_{eq,p}} { N_{0,p} } \simeq 1
\end{equation}

During the whole irradiation time, when a new species is being formed from the radiolysis of its parental species, a fraction of them is also being dissociated and possibly being sputtered from the ice. This last effect is not being considered in this model. Since not all the parental molecules are converted into a given species, some could be sputtered from the ice and others can react to produce different daughter species, the amount of parental molecule always decreases. Consequently, the column density of a given daughter species decreases after it reaches a maximum (equilibrium stage). This behavior can be observed in the case of the evolution of CO$_3$ species as a function of fluence (Fig.~\ref{fig:NiNp}b).

Calling the maximum amount of newly formed species $i$ at the equilibrium as $N_{\infty,i}$ and using Eq.\ref{eq:SIG2}, we can be express the constant $B_i$ of Eq.~\ref{eq:SigmaForm} by:

\begin{equation}\label{eq:SIG3}
B_i = \frac{N_{eq,i}} { N_{0,p}} \simeq  1 - \frac{N_{eq,p}} { N_{0,p} }
\end{equation}

Moreover, considering that the number of initial daughter species $i$ is negligible ($A_i=0$) and, after the combination of Eq.~\ref{eq:SIG2} with Eq.~\ref{eq:SIG3}, we rewrite Eq.~\ref{eq:SigmaForm} as:

\begin{equation}\label{eq:SigmaNEW}
N_i/N_{0,p} \simeq \frac{1}{1+\frac{\sigma_{d,i}}{\sigma_{f,i}}}  [1- \exp(-\sigma_{d,i} F)]
\end{equation}
where $\sigma_{d,i}$ and $\sigma_{f,i}$ are the dissociation and formation cross section of species $i$.

The lines observed in the Fig.~\ref{fig:NiNp}a and Fig.~\ref{fig:NiNp}b were obtained by fitting the experimental data with Eq.\ref{eq:SigmaNEW}. The proposed model is valid from the beginning of irradiation up to the fluence in which the column density reaches its maximum value. This indicates the equilibrium point between formation and destruction promoted by radiolysis. The formation and dissociation cross section as well respective radiolysis yield determined for the newly formed species are given in Table~\ref{tab:SIG-formation}. We observe that CO presents the highest formation cross section among the studied newly formed species.

\subsubsection{Hydrogen peroxide, H$_2$O$_2$}

In recent years, different groups have studied the formation of hydrogen peroxide by ion bombardment of pure water ice and mixed water-CO$_2$ ices (Moore \& Hudson 2000, Strazzulla et al. 2003, 2005a, 2005b; Baragiola et al. 2004; Loeffler et al., 2006; Gomis et al. 2004a, 2004b). H$_2$O$_2$ was also observed from fast electron bombardment of pure water ices (Baragiola, et al. 2005; Zheng et al. 2006) and also from low energy (3-19 eV) electron bombardment (Pan et al. 2004).

Following Teolis et al. (2009), the most commonly H$_2$O$_2$ formation mechanisms involves the reaction between two OH radicals coming from the radiolysis of water molecules, as given by:
\begin{equation}
H_2O \stackrel{CR}{\longrightarrow} H_2O^*{\longrightarrow} OH + H
\end{equation}

\begin{equation}
OH + OH \rightarrow H_2O_2
\end{equation}
where CR denotes cosmic rays.

These OH radicals are strongly hydrogen bonded to water molecules (Cooper et al., 2003), and it is not until above 80 K that they can diffuse within a water-ice lattice (Johnson and Quickenden, 1997). As pointed out by Cooper et al. (2008) energetic OH radicals can diffuse short distances along ion tracks and react at 10 K, but bulk diffusion probably does not occur.

The decrease in the H$_2$O$_2$ production with temperature increase was investigated by Zheng et al. (2006) and Loeffler et al. (2006). This indicates that electron scavenging may play a critical role in the radiation stability of H$_2$O$_2$ in pure water-ice experiments. Moreover, Gomis et al. (2004a) found that the H$_2$O$_2$ yields were dependent on projectile: ices irradiated with low energy (30 keV) C$^+$, H$^+$ and O$^+$ ions produced more H$_2$O$_2$ at 77 K than 16 K, while N$^+$ and Ar$^+$ had no temperature dependence on the H$_2$O$_2$ yield. The authors also observe that the H irradiation produces a much lower quantity of H$_2$O$_2$ than for the other heavy ions suggesting that the energy deposited by elastic collisions plays an important role in such a process.

\begin{table*}[!t]
\begin{center} %
\caption{Comparison between the formation and dissociation cross section of H$_2$O$_2$ from the processing of pure H$_2$O ices and H$_2$O:CO$_2$ ices. The formation and dissociation radiochemical yield of H$_2$O$_2$ as well the asymptotic H$_2$O$_2$/H$_2$O ratio in each experiment are also given.}  \label{tab:COMPAR-H2O2}
\setlength{\tabcolsep}{5pt}
\begin{tabular}{ l l l c c c c c r}
\hline \hline
Ices & Temp. & Projectile  & $\sigma_f$          & $\sigma_d$         & G$_f$             &-G$_d$             &  H$_2$O$_2$/H$_2$O$^\dag$  & Ref. \\
     & (K)   & (Energy)    & $(10^{-13}$ cm$^2$) & $(10^{-13}$ cm$^2$)& (molec / 100 eV)  & (molec / 100 eV)  &   (\%)                     &  \\
\hline
H$_2$O      & 13 & Ni$^{13+}$ (52 MeV)           & 0.19  & 10    & 1.3  & 68.0  & 1.9  & [1]\\
H$_2$O:CO$_2$ (10:1) & 13 & Ni$^{13+}$ (52 MeV)  & 0.13  & 9.3   & 0.87 & 59.1  & 1.6  & [1]\\
H$_2$O:CO$_2$ (1:1) & 13 & Ni$^{13+}$ (52 MeV)   & 0.10  & 6.9   & 0.48 & 33.8  & 1.5  & [1] \\
H$_2$O      & 16 & H$^+$ (30 keV)         & $\sim$ 0.001 & 0.06  & 0.45  & 271  & 0.21  & [2] \\
H$_2$O      & 16 & C$^+$ (30 keV)         & $\sim$ 0.001 & 0.07  &  0.23 & 15.8 & 1.7  & [2] \\
H$_2$O      & 16 & N$^+$ (30 keV)         & $\sim$ 0.001 & 0.09  &  0.19 & 17.1 & 1.6  & [2] \\
H$_2$O      & 16 & O$^+$ (30 keV)         & 0.002        & 0.13  &  0.34 & 22.3 & 1.4  & [2] \\
H$_2$O      & 16 & Ar$^+$ (30 keV)        & 0.002        & 0.07  &  1.34 & 46.8 & 2.2  & [2] \\
H$_2$O      & 20 & H$^+$ (100 keV)      & $\sim$ 0.001   & 0.20  &  0.38 & 76.1 & 0.75  & [3] \\
H$_2$O      & 16 & H$^+$ (200 keV)      & $\sim$ 0.001   & 0.09  &  0.49 & 43.9 & 1.2  & [4] \\
H$_2$O      & 16 & He$^+$ (200 keV)     &          0.003 & 0.13  &  0.37 & 16.1 & 2.4  & [4] \\
H$_2$O & 16 & Ar$^{2+}$ (400 keV)       & $\sim$0.001    & 0.04  &  0.07 & 2.79 & 6.0  & [4] \\
\hline \hline \\
\end{tabular}
\end{center}
\vspace{-0.5cm}
$^\dag$ Asymptotic value; [1] this study; [2] Gomis et al. 2004a; [3] Loeffler et al. 2006; [4] Gomis et al. 2004b.
\end{table*}

As discussed by Zheng et al. (2006) and Loeffler et al. (2006), there is another reaction sequence to produce hydrogen peroxide in astrophysical ices from the addition of oxygen atom (a product of the by the dissociation of oxidant compounds such as O$_2$, CO, CO$_2$, H$_2$O, etc.) to water molecule:
\begin{equation}
\textrm{O} + \textrm{H$_2$O} \rightarrow \textrm{H$_2$O-O} \stackrel{ionization}{\longrightarrow} \textrm{H$_2$O$^+$ + O$^-$}  \rightarrow \textrm{H$_2$O$_2$}
\end{equation}
This reaction route involves an extra ionization stage triggered by CR, UV/X-ray photons or fast electrons to produce H$_2$O$_2$ (Loeffler et al. 2006), which implies a quadratic dependence on irradiation fluence being a secondary order reaction processes. In the present work we consider negligible the H$_2$O$_2$ formation via oxygen addition to H$_2$O. Future investigations with isotopic labeling could help to clarify this issue and quantify which fraction of O atoms produced from the dissociation of CO$_2$ or H$_2$O may react with H$_2$O to form H$_2$O$_2$.

Table~\ref{tab:COMPAR-H2O2} presents the formation and dissociation cross section, as well the radiochemical yield, of H$_2$O$_2$ obtained from the processing of pure water ice and mixed H$_2$O:CO$_2$ ice. Both the formation and dissociation cross section of H$_2$O$_2$ decrease as the relative abundance of H$_2$O in the ice decreases, a consequence of larger averaged distance between parental species. From Fig.~\ref{fig:NiNp} we also observe that the presence of CO$_2$ inside the ice decreases the H$_2$O$_2$/H$_2$O ratio. This points in the opposite direction than the results observed from the irradiation of water-CO$_2$ by light ions (Strazzulla et al. 2005b; Moore and Hudson 2000). From Table~\ref{tab:COMPAR-H2O2} we observe that the formation cross section of H$_2$O$_2$ from the bombardment of pure water ice by heavy and energetic ions is 10$^5$ higher than the value obtained by the impact of light or/and slower projectiles.

The asymptotic value for the H$_2$O$_2$/H$_2$O ratio presents a slight reduction with the enhancement of initial CO$_2$ in the ice. This clearly shows that H$_2$O$_2$ formation via O + H$_2$O in which the oxygen comes from another oxidant compound, such as CO$_2$, is indeed a secondary-order process (Loeffler et al. 2006). Our value is about 8 times higher than in the case of irradiation whit 30 keV protons, but very similar to the asymptotic values obtained after the impact with other 30 keV ions such as C$^+$, N$^+$ and O$^+$ (Gomis et al 2004a). A comparison between our radiochemical yield and cross sections with literature have revealed higher values for both the formation and destruction of H$_2$O$_2$ during bombardments of ices by heavy and energetic ions. In other words the chemical processing is very enhanced by the processing of ices by heavy and energetic and highly charged ions in comparison with light ions.

Hydrogen peroxide has been found on the surface of Europa by identifying both an absorption feature at 3.5 $\mu$m in the Galileo NIMS spectra and looking at the UV spectrum taken by the Galileo ultraviolet spectrometer (UVS) (Carlson et al. 1999). Following the authors, the relative abundance with respect to water in Europa is H$_2$O$_2$/H$_2$O $\sim$ 0.001 ($=$0.1\%). As the Jupiter satellite is immersed into the intense magnetosphere of the planet, the authors also suggest that radiolysis is the dominant formation mechanism of such a molecule. As discussed by Hendrix et al. (1999) this compound has also been detected on the other icy Galilean moons of Jupiter in the ultraviolet, such as Callisto and Ganymede, although it is less evident on the darker satellites than on Europa. The estimated ratio on Ganymede is  H$_2$O$_2$/H$_2$O $\sim$ 0.003 ($=$0.3\%) (Hendrix et al. 1999).

H$_2$O$_2$ has also been suggested to be present in icy mantles on grains in dense clouds as pointed out Boudin et al. (1998). Based on observations and laboratory studies, they have estimated the relative abundance with respect to water at the dense molecular cloud NGC 7538:IRS 9 of H$_2$O$_2$/H$_2$O $\sim$ 0.05 ($=$5\%). Following the authors, hydrogen peroxide can be also produced on grain surfaces via the hydrogenation of molecular oxygen.

\subsubsection{Ozone, O$_3$}

Ozone has been detected after irradiation of mixed H$_2$O:CO$_2$ (1:1) ices  and pure CO$_2$ ice at 16 K by 1.5-200 keV ions (protons and He$^+$) (Strazzulla et al. 2005a, 2005b). According to these authors, the O$_3$ production from radiolysis is highly attenuated in the case of CO$_2$-poor mixed ices, e.g. for H$_2$O/CO $>$ 2.5. We do not observe ozone among the compounds produced by radiolysis of H$_2$:CO$_2$ (10:1) ice. The presence of ozone has also been observed after the irradiation of CO:O$_2$ (1:1) ice by 60 keV Ar$^{++}$ and 3 keV He$^+$ (Strazzulla et al 1997). The authors suggest that the detection of ozone and the absence of suboxides (e.g. C$_2$O and C$_3$O) in interstellar grains would indicate a dominance of molecular oxygen in grain mantles. These suboxides have not been detected in our experiments.

Seperuelo Duarte et al. (2010) have bombarded pure CO ice with 50 MeV Ni$^{+13}$ and observe ozone and several suboxides (C$_2$O, C$_3$O$_2$, C$_5$O$_2$) among the radiolysis products. They determined the ozone formation cross section is 3 $\times 10^{-16}$~cm$^{2}$. Ozone and suboxides were also observed after irradiation of pure CO ice at 16 K by 200 keV protons (Palumbo et al. 2008).

We have detected ozone after the radiolysis of pure CO$_2$ ice and mixed H$_2$O:CO$_2$ (1:1) ice by 52 MeV Ni$^{+13}$. The determined formation cross section of ozone is $\sigma_{O3} = 0.27~\times~10^{-13}$ cm$^{2}$ and $0.06 \times 10^{-13}$ cm$^{2}$ in pure CO$_2$ ice and  H$_2$O:CO$_2$ ice, respectively. These values indicate that the O$_3$ formation cross section decreases as the relative abundance of the parental CO$_2$ compound in the ice declines. The value obtained for ozone formation via pure CO$_2$ is 2 times higher than the value determined previously in similar radiolysis experiments involving pure $^{18}$CO$_2$ irradiated by 46 MeV 58 Ni$^{+11}$ (Seperuelo Duarte et al. 2009).

In the solar system, ozone has been observed on Ganymede (Noll et al. 1996) and on Saturn's satellites Rhea and Dione (Noll et al. 1997). Following Strazzulla et al. (2005a) ozone could be formed where fresh CO$_2$ rich layers are exposed to radiation. Due to the strong silicate absorption band around 1040 cm$^{-1}$ the observation of ozone in interstellar medium is very difficult. However, the appearance of ozone seems conceivable, since it is observed in laboratory experiments involving the processing of astrophysical ices analogs.

\subsubsection{CO$_3$}

This compound has been observed after the UV processing of pure CO$_2$, CO and O$_2$ 10 K ices (Gerakines et al. 1996; Gerakines \& Moore 2001) and also on radiolysis of pure CO$_2$ by 0.8 MeV protons (Gerakines \& Moore 2001) and 1.5 keV protons (Brucato et al. 1997). The bombardment of mixed H$_2$O:CO$_2$ ices by 3 keV He$^+$, 1.5 keV and 0.8 MeV protons also shows the formation of CO$_3$. Furthermore, H$_2$CO$_3$ and possibly O$_3$ are produced (Brucato et al. 1997; Moore and Khanna 1991).

Seperuelo Duarte et al. (2009) have observed this species in the radiolysis of
of C$^{18}$O$_2$ by 46 MeV $^{58}$Ni$^{11+}$ up to a final fluence of $1.5 \times 10^{13}$ cm$^{-2}$. In our measurement the formation cross section of CO$_3$ from radiolysis of pure CO$_2$ ice is $0.08 \times 10^{-13}$ cm$^{-2}$, roughly 5 times lower than the value obtained previously by Seperuelo Duarte et al. (2009). This value presents a decrease in a presence of water in the ice, as observed in the case of H$_2$O:CO$_2$ (1:1) ice (see Table~\ref{tab:SIG-formation}). The formation cross section of CO in our experiments on pure CO$_2$ is roughly the same than the one determined by Seperuelo Duarte et al. (2009).

The CO$_3$ destruction cross section due to radiolysis is very large when compared to the other compounds formed into the ices. To increase the accuracy of the cross sections next measurements should cover large data sets inside the fluence range between 1-10$\times 10^{11}$ ions cm$^{-2}$. On the contrary of UV photolysis of pure CO$_2$ ices (Gerakines et al. 1996) we do not observed the formation of C$_3$O at 2243 cm$^{-1}$ from the radiolysis of CO$_2$ ices.

Despite of the prediction of CO$_3$ among the compounds in the processed astrophysical ices (e.g. Elsila et al. 1997, Allamandola et al. 1999), this species has not been detected conclusively in space yet (Elsila et al. 1997, Ferrante et al. 2008).

\subsubsection{H$_2$CO$_3$}

Carbonic acid (H$_2$CO$_3$) has been observed in several experiments involving ion bombardment of  H$_2$O:CO$_2$ ices (Moore \& Khanna, 1991; DelloRusso et al. 1993; Brucato et al. 1997). It was also observed after the radiolysis of pure CO$_2$ ice at 10 K by 1.5 keV protons (Brucatto et al. 1997), indicating that implanted hydrogen ions are incorporated in the target to form new bonds to produce H$_2$CO$_3$.

As discussed by Gerakines et al. (2000) the formation of H$_2$CO$_3$ from ion bombardment of H$_2$O:CO$_2$ ices is ruled by two main reaction schemes. First, the direct dissociation of H$_2$O molecules by incoming projectile:
\begin{equation}
\textrm{H$_2$O} \stackrel{CR}{\longrightarrow} \textrm{OH} + \textrm{H$^+$} + e^-
\end{equation}
where products such as H$_2$, H$_2$O$_2$; H$_3$O$^+$, and HO$_2$ are eventually formed by reactions involving the primary dissociation products. The next step is the electron attachment on CO$_2$ or OH producing reactive compounds which quickly react with each other to produce bicarbonate (HCO$_3^-$):
\begin{equation}
\textrm{CO$_2^-$} + \textrm{OH}  {\longrightarrow} \textrm{HCO$_3^-$}
\quad
\textrm{and/or}
\quad
\textrm{CO$_2$} + \textrm{OH$^-$}  {\longrightarrow} \textrm{HCO$_3^-$}
\end{equation}
Finally, bicarbonate reacts with a proton to produce H$_2$CO$_3$:
\begin{equation}
\textrm{HCO$_3^-$} + \textrm{H$^+$}  {\longrightarrow} \textrm{H$_2$CO$_3$}
\end{equation}

From the evolution of the IR peak at 1307 cm$^{-1}$ during the radiolysis of H$_2$O:CO$_2$ (1:1) ice by 52 MeV Ni ions we have determined the formation and the dissociation cross section of H$_2$CO$_3$ and the values obtained are $3.0 \times 10^{-15}$ cm$^{2}$ and $9.9 \times 10^{-13}$ cm$^{2}$, respectively. This formation cross section is 8 orders of magnitude higher than the one derived from the UV photolysis of H$_2$O:CO (1:1) ices at 18 K (Gerakines et al. 2000) indicating that heavy ion processing of astrophysical ice analogs is very efficient to form H$_2$CO$_3$ compared to UV photons.

H$_2$CO$_3$ is thermally stable at 200 K, higher than the sublimation temperature of H$_2$O. As discussed by Peeters et al. (2008), ices containing both H$_2$O and CO$_2$ have been found on a variety of surfaces such as those of Europa, Callisto, Iapetus, and Mars, where processing by magnetospheric ions or the solar wind and energetic solar particles may lead to the formation of H$_2$CO$_3$.

The astrophysical significance of solid carbonic acid have been extensively discussed elsewhere (e.g. Khanna et al. 1994, Brucato et al 1997, Strazzullla et al. 1996). H$_2$CO$_3$ has been suggested to be present at the surface of several solar system bodies with sufficient amounts of CO$_2$ such as Galilean satellites of Jupiter (Wayne 1995, McCord et al. 1997, Johnson et al. 2004) and comets (Hage et al. 1998). Europa, Ganymede, and Callisto all exhibit high surface CO$_2$ abundances and these satellites are heavily bombarded by energetic magnetospheric particles, galactic cosmic rays and solar radiation.

It has been suggested that the 3.8 $\mu$m feature in NIMS spectra of Ganymede and Callisto arise from H$_2$CO$_3$ (Johnson et al. 2004; Hage et al. 1998). Following Carlson et al. (2005) the relative abundance of carbonic acid with respect to CO$_2$ in Callisto is H$_2$CO$_3$/CO$_2$ $\sim$ 0.01.

The presence of carbonic acid in interstellar ices seems also to be very likely (Whittet et al. 1996) not only in the solid phase but also in gas phase (Hage et al. 1998). In the case of the quiescent cloud medium toward the background field star Elias 16, the expected H$_2$CO$_3$ column density could be as much as $4.6 \times 10^{16}$ molec cm$^{-2}$, corresponding roughly 1\% of the CO$_2$ or 0.2\% of the total H$_2$O in this line of sight (Whittet et al. 1998).

As discussed by Zheng \& Kaiser (2007), in the history of Mars there might be a high concentration of carbonic acid produced by radiolysis of surface H$_2$O:CO$_2$ ices. Without a dense atmosphere and magnetic field, Mars lacks the power to attenuate penetrating energetic solar particles, energetic cosmic rays or any type of high energy particles. Following Westall et al. (1998), if available in sufficient concentrations, carbonic acid could potentially dissolve metal ores and catalyze chemical reactions and its presence on Mars may also lead to the existence of limestone (CaCO$_3$), magnesite (MgCO$_3$), dolomite (CaMg(CO$_3$)$_2$) and siderite (FeCO$_3$).

\section{Astrophysical implication}

\begin{figure}[t]
 \centering
\includegraphics[scale=0.72]{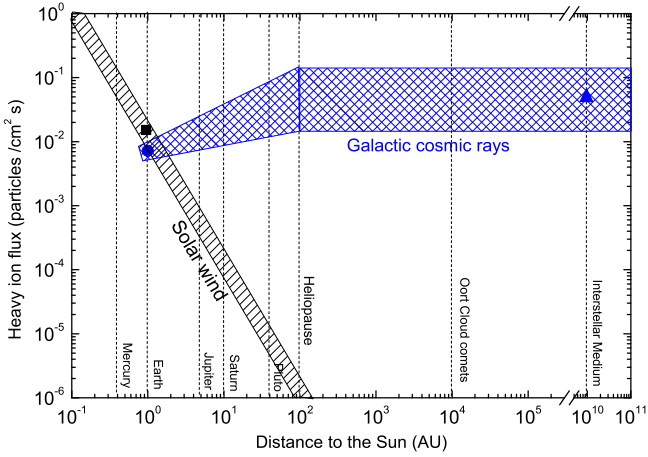}
 \caption{Estimated value of the integrated flux of heavy ions (12~$\lesssim$~Z~$\lesssim$~29) with energy between 0.1-10 MeV/u inside solar system and at interstellar medium as a function of distance to the Sun. Both Galactic cosmic rays and energetic solar particles are displayed. Square: integrated flux of energetic solar particles. Circle and triangle: integrated flux of cosmic rays.} \label{fig:FLUX}
\end{figure}

Figure~\ref{fig:FLUX} presents the total flux of heavy ions (12~$\lesssim$~Z~$\lesssim$~29) with energy between 0.1-10 MeV/u inside solar system as a function of distance to the Sun. Both Galactic cosmic rays (GCR) and energetic solar particles are displayed. The integrated flux of heavy and energetic solar ions (black square) at Earth Orbit were measured by Mewaldt et al. (2007). Starting from the solar photosphere abundance and the asumtion that elements with first ionization potential of less than 10 eV are more abundant in the energetic solar particles than in photosphere by a factor of 4.5 (Grevesse et al. 1996) at Earth orbit, the integrated flux of heavy and energetic solar particles ($\phi_{HSW}$) with energies around 0.1-10 MeV/u was found to be about 1.4 $\times$ 10$^{-2}$ cm$^{-2}$ s$^{-1}$ (black square in Fig.~\ref{fig:FLUX}).

The heavy ion integrated flux at Earth orbit from galactic sources (heavy cosmic rays) was calculated by using the lunar-GCR particle model of Reedy and Arnold (1972) (see also Fig. 3.20 of Vaniman et al. 1991) and taking into an account the relative elemental abundances of $12<Z<29$ atoms at 1 AU (Simpson 1983; Drury, Meyer, Ellison 1999). The value obtained was $\sim 7 \times 10^{-3}$~cm$^{-2}$~s$^{-1}$ (blue circle in Fig.~\ref{fig:FLUX}). Assuming that the fluxes of solar wind and energetic solar particles are inversely proportional to the squared distance, this value can be determined as a function of distance up to the heliopause ($\sim$100 AU) (i.e. the boundary between the solar wind domain and the interstellar medium).

The estimation of the average heavy nuclei cosmic ray flux ($\phi_{HCR}$) in interstellar medium was performed by (Pilling et al. 2010), which considered the value of $\phi_{HCR} \sim 5 \times 10^{-2}$~cm$^{-2}$~s$^{-1}$ (blue triangle in Fig.~\ref{fig:FLUX}). This value was rather the same as at the outer border of heliopause. An error region (dashed area) was introduced in the Fig.~\ref{fig:FLUX} to take into an account the uncertainty of these estimations.

The value of integrated flux of heavy ions with energies around 0.1-10 MeV/u from galactic sources and energetic solar particles are comparable at Mars orbit ($\sim$1.5 AU). However at the orbits of Jupiter, Saturn, Uranus, Pluto and at the heliopause border, $\phi_{HCR}$/$\phi_{HSW} \sim$ 2 $\times$ 10$^1$, 1 $\times$ 10$^2$, 7 $\times$ 10$^2$, 3 $\times 10^3$ and 2 $\times 10^4$, respectively. The total flux of heavy particles with energy between 0.1-10 MeV/u at Oort cloud distance was assumed to be the same as expected in the interstellar medium.

Considering the estimated flux of heavy particles of energetic solar particles ($\phi_{HSW}$) and of heavy nuclei cosmic rays ($\phi_{GCR}$) as well the determined dissociation cross section ($\sigma_d$), we can calculate the typical molecular half-lives of frozen molecules in astrophysical surfaces due the presence of heavy particles by the expression (see Pilling et al. 2010):

\begin{equation}
t_{1/2} \approx \frac{\ln 2}{(\phi_{HSW} + \phi_{HCR}) \times \sigma_d} \quad \textrm{[s$^{-1}$]}
\end{equation}

\begin{figure}[!t]
 \centering
\includegraphics[scale=0.40]{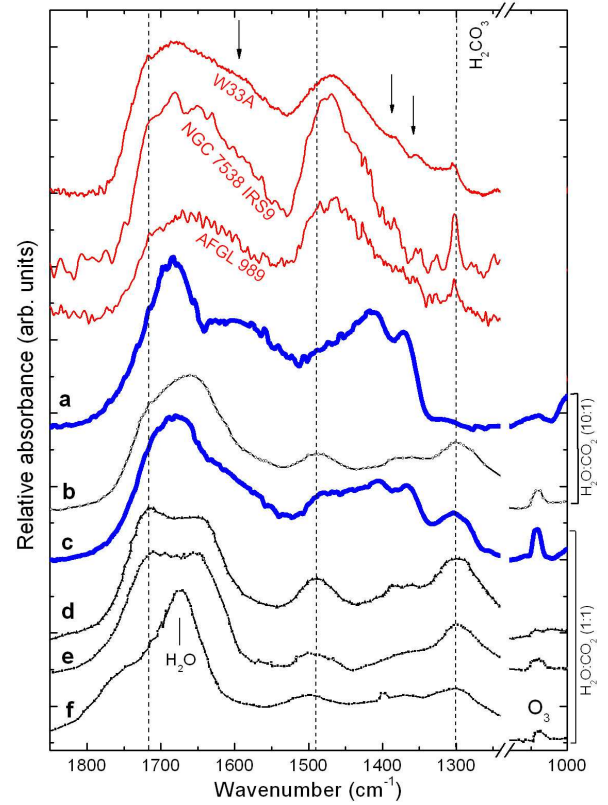}
 \caption{Comparison between IR spectra of interstellar and laboratory
ices. The top three curves are infrared spectra of young stellar sources obtained by the Infrared Space Observatory (ISO). Lower traces indicate different laboratory spectra of irradiated H$_2$O:CO$_2$ ices: a and c (this work); b (Hudson \& Moore 2001); d (Gerakines et al. 2000); e (Strazzulla et al. 2005b); f (UV photons, Gerakines et al. 2000). Vertical dashed lines indicate the location of vibrational modes of frozen H$_2$CO$_3$.} \label{fig:comp}
\end{figure}

It is worth to note that the present experiments were performed at low temperature of 13 K, the temperature that is adequate to ices in the interstellar medium. Therefore it is useful to make a comparison between IR spectra of interstellar and laboratory ices is shown in Figure~\ref{fig:comp}. The top three curves are IR spectra of young stellar sources obtained by the Infrared Space Observatory (ISO). The two bold (blue) curves are our data. The six bottom curves indicate different laboratory spectra of processed H$_2$O:CO$_2$ ices: a) mixture (10:1) irradiated by 52 MeV Ni ions (this work); b) mixture (10:1) irradiated by 0.8 MeV protons (Hudson \& Moore 2001); c) mixture (1:1) irradiated by 52 MeV Ni ions (this work); d) mixture (1:1) irradiated by 0.8 MeV protons (Gerakines et al. 2000); e) mixture (1:1) irradiated by 0.8 MeV protons (Strazzulla et al. 2005b) and f) mixture (1:1) irradiated by UV photons (Gerakines et al. 2000). Vertical dashed lines indicate the location of vibrational modes of frozen H$_2$CO$_3$ (Moore \& Khanna 1991). The presence of H$_2$CO$_3$ seems to be independent of the ionization source.

The three bumps observed on the IR spectra of the young stelar objects at about 1600~cm$^{-1}$, 1400~cm$^{-1}$ and 1350~cm$^{-1}$ (see arrows on Fig~\ref{fig:comp}) present a good similarity with the features present in the IR spectra of H$_2$O:CO$_2$ (10:1) irradiated by 52 MeV ions (curve a). Nevertheless, up to know, its molecular assignment remains unknown. Although very attenuated, these features are still observed in the IR spectra of H$_2$O:CO$_2$ (1:1) (curve c). This suggests that heavy cosmic rays can be a good candidate to explain some features observed in the IR spectra of some interstellar sources.

The comparison between the current results with the previous one (Pilling et al. 2010) reveals that, independently of ice constitution (involving H$_2$O, CO$_2$, CO and NH$_3$), the dissociation cross section due to heavy and energetic cosmic ray analogs are in the same range, $\sigma_d=$~1-2~$\times 10^{-13}$ cm$^{2}$. This supports the extension of our previous estimative for the half-life of ammonia-containing ice, $\tau_{1/2} =$~2-3~$\times$ 10$^6$ years, to all kind interstellar grains inside of dense interstellar environments in the presence of a constant galactic heavy cosmic ray flux.

The temperature of solar system ices is about 80 K, about 5-8 times higher than observed in interstellar ices, therefore some molecular species (the most volatile as O$_3$ and CO) are not efficiently trapped/adsorbed in/on the ices. This issue can makes an enormous difference on surface and bulk chemistry. Future experiments employing H$_2$O:CO$_2$ ices at 80 K will be performed to investigate this issue.

\section{Summary and conclusions}

The interaction of heavy, highly charged and energetic ions (52 MeV $^{58}$Ni$^{13+}$) with pure H$_2$O and CO$_2$ ices and mixed (H$_2$O:CO$_2$) ices was studied experimentally. The aim was to understand the chemical and the physicochemical processes induced by heavy cosmic rays inside dense and cool astrophysical environments such as molecular clouds, protostellar clouds as well at the surfaces solar system ices. Our main results and conclusions are the following:
\begin{enumerate}
  \item In all experiments containing CO$_2$ (pure ice and mixtures) after a fluence of about 2-3 $\times 10^{12}$ ions cm$^{-2}$, the CO$_2$/CO ratio became roughly constant ($\sim$ 0.1), independently on initial CO$_2$/H$_2$O. The experiments suggest that the abundances of CO$_3$, O$_3$, H$_2$CO$_3$ in typical astrophysical ices ([CO$_2$]/[H$_2$O] $\sim$ 0.05-0.1) irradiated by heavy ions, should be very low, except for those ices with peculiar enrichment of CO$_2$.

  \item After a fluence of about 2-3 $\times 10^{12}$ ions cm$^{-2}$ the H$_2$O$_2$/H$_2$O ratio stabilizes at $\sim$ 0.01, not depending of initial H$_2$O relative abundance and column density.

  \item A comparison between our radiochemical yield and cross sections with literature have revealed higher values for both the formation and destruction of H$_2$O$_2$ during bombardments of ices by heavy and energetic ions. In other words the chemical processing is very enhanced by the processing of ices by heavy and energetic and highly charged ions in comparison with light ions.

  \item The dissociation cross section of pure H$_2$O and CO$_2$ ices are 1.1 and 1.9 $\times 10^{-13}$ cm$^2$, respectively. In the case of mixed H$_2$O:CO$_2$ (10:1) the dissociation cross section of both species is roughly 1 $\times 10^{-13}$ cm$^2$.

 \item Some IR features observed in ices after the bombardment by heavy ions present great similarity with those observed at molecular clouds, suggesting that heavy cosmic rays play an important role in the processing of frozen compounds in interstellar environments.

\end{enumerate}

%
\begin{acknowledgements} The authors acknowledge the agencies COFECUB (France) as well as CAPES, CNPq and FAPERJ (Brazil) for partial support. We thank Th. Been, I. Monnet, Y. Ngono-Ravache and J.M. Ramillon for technical support.

\end{acknowledgements}
%
%


\begin{thebibliography}{}

\bibitem{}
Allamandola, L.J., Bernstein M.P., Sandford S.A., Walker R.L., 1999, Space Science Reviews 90, 219

\bibitem{}
Baragiola R.A., Loeffler M.J., Raut U., Vidal R.A., Carlson R.W., 2004, Lunar and Planetary Science Conference, 35, 2079

\bibitem{}
Baragiola R.L.A., Loeffler M.J., Raut U., Vidal R.A., \& Wilson C.D., 2005, Radiation Physics and Chemistry, 72, 187

\bibitem{}
Bennett C.J., Jamieson C.S., Mebel A.M., \& Kaiser R.I. 2004, Phys.
Chem. Chem. Phys., 6, 735

\bibitem{}
Bockelée-Morvan D., 1997, In Molecules in
Astrophysics: Probes and Processes, pg. 218, van Dishoeck E.F. (ed.), Kluwer, Dordrecht

\bibitem{}
Boudin N., Schutte W.A., and Greenberg J.M. 1998, A\&A, 331, 749

\bibitem{}
Bouwman J., Ludwig W., Awad Z., Öberg K.I., Fuchs G.W., van Dishoeck E.F. \& Linnartz H., 2007, A\&A, 476, 995

\bibitem{}
Brown W.L., Augustyniak W.M., Marcantonio K.J., Simmons E.H., Boring J.W., et al., 1984, Nucl. Instr. and Meth. B1, IV, 307

\bibitem{}
Brucato J.R., Palumbo M.E., Strazzulla G., 1997, Icarus 125, 135

\bibitem{}
Carlson R.W., Anderson M.S., Johnson R.E. et al., 1999, Science, 283, 2062

\bibitem{}
Carlson R.W., Hand K.P., Gerakines P.A., Moore M.H. \& Hudson R.L. 2005, Bulletin of the American Astronomical Society, 37, 5805

\bibitem{}
Cooper P.D., Moore M.H., Hudson R.L. et al., 2008, Icaurs, 194, 379

\bibitem{}
Cooper P.D., Johnson R.E. \& Quickenden T.I., 2003, Icarus 166, 444

\bibitem{}
Crovisier J. et al.,  1997, Science, 275, 1904.

\bibitem{}
Crovisier J., 1998, Faraday Discuss., 109, 437.

\bibitem{}
DelloRusso et al. 1993, JGR, 98, 5505

\bibitem{}
Drury L.O.C., Meyer J.-P., Ellison D.C., 1999, ``Topics in Cosmic-Ray Astrophysics", M.A. DuVernois ed., Nova Science Publishers, New-York

\bibitem{}
Ehrenfreund P. \& Charnley S.B. 2000, ARAA, 38, 427

\bibitem{}
Elsila J., Allamandola L.J. and Sandford S.A., 1997, ApJ, 479, 818

\bibitem{}
Ferrante R.F., Moore M.H., Morgan M., Spiliotis M.M. and Hudson R.L.. 2008, ApJ, 684, 1210

\bibitem{}
Gerakines P.A., Moore M.H. \& Hudson R.L. 2000 A\&A, 357, 793

\bibitem{}
Gerakines P.A., Whittet D.C.B., Ehrenfreund, P., et al., 1999, ApJ, 522, 357

\bibitem{}
Gerakines P.A. et al. 1996, A\&A, 312, 289

\bibitem{}
Gerakines P.A. \& Moore M.H. 2001, Icarus, 154, 372

\bibitem{}
Gerakines P.A., Scutte W.A., Greenberg J.M., van Dishoeck E. F., 1995, A\&A, 296, 810.

\bibitem{}
Gibb E.L., Whittet D.C.B., Schutte W.A., et al., 2000, ApJ, 536, 347.

\bibitem{}
Gomis et al. 2004a, PSS, 52, 371

\bibitem{}
Gomis et al. 2004b, A\&A, 420, 405.

\bibitem{}
Grevesse N., Noels A. \& Sauval J. A., 1996, ASP Conference Series, 99, 117

\bibitem{}
Hage W., Liedl K.R., Hallbrucker A., Mayer E., 1998, Science, 279, 1332

\bibitem{}
d'Hendecourt L.B., \& Allamandola L.J., 1986, A\&ASS, 64, 453

\bibitem{}
Hendrix A.R., Barth C.A., Stewart A.I.F., Hord C.W., Lane A.L. 1999, LPI, 30,2043


\bibitem{}
Herr K.C. \& Pimentel G.C., 1969, Science 166, 496

\bibitem{}
Hudson R.L. \& Moore M.H., 2001, J. Geoph. Res., 106, 33275

\bibitem{}
Iza P., Farenzena  L.S., Jalowy T., Groeneveld K.O. \& E.F. da Silveira, 2006, Nuclear Instruments and Methods in Physics Research Section B, 245, 61

\bibitem{}
Johnson R. E. \& Quickenden T.I., 1997, JGR, 102, 10985J

\bibitem{}
Johnson R.E., Carlson R.W., Cooper J.F., Paranicas C., Moore M.H. and Wong M.C., 2004, Radiation effects on the surfaces of the Galilean satellites, In F. Bagenal,
T.E. Dowling, and W.B. McKinnon, Eds., Jupiter: The planet, satellites and magnetosphere, vol. 1, pages 485-512, Cambridge University Press.

\bibitem{}
Khanna R. K. \& Tossell J.A., Fox F., 1994, Icarus, 112, 541

\bibitem{}
Larson H.P. \& Fink U. 1972, ApJ 171, L91

\bibitem{}
Leto G. \& Baratta G.A., A\&A, 397, 7

\bibitem{}
Loeffler M.J., Raut U., Vidal R.A., Baragiola R.A., Carlson R.W., 2006, Icarus, 180, 265

\bibitem{}
Loeffler M.J., Barata G. A., Palumbo M.E., Strazulla G. and Baragiola R.A., A\&A, 2005, 435, 587

\bibitem{}
McCord T.B., Hansen G.B., Clark R.N. et al., 1998, J. Geophys. Res.
103, 8603

\bibitem{}
McCord T.B., et al., 1997, Science, 278, 271

\bibitem{}
Moore M. \& Hudson R. L., 2000, Icarus, 145, 282


\bibitem{}
Moore M.H. \& Khanna R.K. 1991, Spectrochim. Acta 47, 255

\bibitem{}
Mewaldt, R. A., Cohen, C. M. S., Mason, G. M., Haggerty, D. K., \& Desai, M. I. 2007, Space Sci. Rev., 130, 323

\bibitem{}
Nastasi M., Mayer J. \& Hirvonen J.K., 1996, in ``Ion-Solid Interactions: Fundamentals and Applications", Cambridge Solid State Science Series, Cambridge University Press

\bibitem{}
Noll, K. S., Johnson, R. E., Lane, A. L., Domingue, D. L., \& Weaver, H. A. 1996, Science 273, 341

\bibitem{}
Noll, K. S., Roush, T. L., Cruikshank, D. P., Johnson, R. E., \& Pendleton, Y. J. 1997, Nature 388, 45

\bibitem{}
Öberg K.I., Fraser H.J., Boogert A.C.A.,  Bisschop S.E., Fuchs G.W., van Dishoeck E. F. \& Linnartz H. 2007, A\&A, 462, 1187

\bibitem{}
Pan, X. N., Bass, A. D., Jay-Gerin, J. P., \& Sanche, L. 2004, Icarus, 172, 521;

\bibitem{}
Palumbo M.E., Leto P., Siringo C. \& Trigilio C., 2008, ApJ, 685, 1033

\bibitem{}
Palumbo M.E., Baratta G.A. \& Spinella F. 2006, Mem. S.A.It. Suppl., 9, 192

\bibitem{}
Palumbo M. E. 2006, A\&A 453, 903

\bibitem{}
Paul J.B., Collier C.P., Saykally R.J.,  Scherer J.J. \& O'Keefe A. 1997, J. Phys. Chem. A, 101, 5211

\bibitem{}
Peeters, Z., Hudson, R., Moore, M., 2008, DPS, 40, 5405

\bibitem{}
Pilling, S.; Seperuelo Duarte, E.; da Silveira, E. F.; Balanzat, E.; Rothard, H.; Domaracka, A.; Boduch, P. 2010, A\&A, 509A, 87

\bibitem{}
Quirico E., Douté S., Schmitt B. et al., 1999, Icarus 139, 159

\bibitem{}
Reedy, R. C.; Arnold, J. R. 1972, J. Geophys. Res., 77, 537

\bibitem{}
Seperuelo Duarte E., Boduch P., Rothard H., Been T., Dartois E., Farenzena L.S. \& da Silveira E. F. 2009, A\&A, 502, 599

\bibitem{}
Seperuelo Duarte E., Domaracka A., Boduch P., Rothard H., Dartois E. \& da Silveira E. F., 2010,  A\&A, 512A, 71.

\bibitem{}
Simpson J. A., 1983, Ann. Rev. Nucl. Part. Sci., 33, 323

\bibitem{}
Smith M.A.H., Rinsland C.P., Fridovich B. \& Rao K.N., 1985, Molecular Spectroscopy: Modern Research, Vol. 3 (London: Academic)

\bibitem{}
Strazzullla G. , Brucato J. R., Cimino G., Palumbo M. E., 1996, Planet. Space Sci., 44, 1447

\bibitem{}
Strazzulla G., Brucato J.R., Palumbo M.E., \& Satorre M.A., 1997, A\&A, 321, 618

\bibitem{}
Strazzulla, G., Leto, G., Gomis, O., and Satorre, M.A., 2003, Icarus 164, 163.

\bibitem{}
Strazzulla G., Leto G., Spinella F. \& Gomis O., 2005a, Astrobiology, 5, 612.

\bibitem{}
Strazzulla G., Leto G., Spinella F. \& Gomis O., 2005b, Mem. S.A.It. Suppl., 6, 51

\bibitem{}
Teolis B.D., Shi J. \& Baragiola R. A., 2009, JCP, 130, 134704

\bibitem{}
Wayne R. P., 1995, Chemistry of the Atmospheres, Chaps. 8 and 9, Clarendon Press, Oxford.

\bibitem{}
Westall F. , Gobbi P. , Gerneke D., Mazzotti G. ,1998, in lunar and planetary institute conference abstracts, p. 1362

\bibitem{}
Whittet et al. 1998, ApJ 498, L159

\bibitem{}
Whittet D.C.B, Schutte W.A., Tielens A.G.G.M., et al., 1996, A\&A, 315, L357

\bibitem{}
Wu C.Y. R. et al. 2003, J. Geophys. Res., 108(E4), 13-1 to 13-8.

\bibitem{}
Vaniman D., Reedy R., Heiken G., Olhoeft G. and Mendell W. 1991 In Lunar Sourcebook: a user's guide to the Moon. Editors: Heiken G.H., Vaniman, D.T., French B.M., pp 55, Cambridge Univ. Press.

\bibitem{}
Zheng W., Jewitt D. \& Kaiser R. 2006, ApJ, 639, 534

\bibitem{}
Zheng W. \&  Kaiser R. I. 2007, Chem. Phys. Lett., 450, 55

\bibitem{}
Ziegler J.F. 2003, Stopping and Range of Ions in Matter SRIM2003 (available at \verb"www.srim.org")

\end{thebibliography}
\end{document}